\documentclass[prb,twocolumn,superscriptaddress,showpacs,letterpaper]{revtex4}

\usepackage{amsmath,amsfonts}
\usepackage{graphicx}
\usepackage{enumerate}

\begin{document}


\title{Spin current and rectification in Luttinger liquids}

\author{Bernd Braunecker}
\affiliation{Department of Physics, Brown University, Providence, RI 02912}
\affiliation{Department of Physics and Astronomy, University of Basel, 
             Klingelbergstrasse 82, CH-4056 Basel, Switzerland}

\author{D. E. Feldman}
\affiliation{Department of Physics, Brown University, Providence, RI 02912}

\author{Feifei Li}
\affiliation{Department of Physics, Brown University, Providence, RI 02912}

\date{\today}




\begin{abstract}
We demonstrate that spin current can be generated by an ac voltage in
a one-channel quantum wire with strong repulsive electron interactions 
in the presence of a non-magnetic impurity and uniform static magnetic field. 
In a certain range of voltages, the spin current can exhibit a power 
dependence on the ac voltage bias with a negative exponent.
The spin current expressed in units of $\hbar/2$ per second can become 
much larger than the charge current in units of the electron charge 
per second. The spin current generation requires neither spin-polarized particle
injection nor time-dependent magnetic fields.
\end{abstract}


\maketitle


\section{Introduction}

The pioneering paper by Christen and B\"{u}ttiker \cite{Christen96} 
has stimulated much interest to rectification in quantum wires and 
other mesoscopic systems. Most attention was focused on the simplest 
case of Fermi liquids \cite{FLRect,magn}. Recently this research was 
extended to strongly interacting systems where Luttinger liquids 
are formed \cite{Feldman05,BB05b,magn-exp,magn-lat}.
One of the topics of current interest is the rectification effect 
in Luttinger liquids in a magnetic field \cite{magn,magn-exp,magn-lat}.
In the presence of a magnetic field, both spin and charge currents 
can be generated. So far, however, only charge currents in Luttinger liquids
have been studied.
In this paper we show that a dc spin current can be generated by an ac
voltage bias in a single-channel quantum wire.

In recent years many approaches to the generation of spin currents in 
quantum wires were put forward. Typically both, spin and charge currents 
are generated and the spin current expressed in units of         
$\hbar/2$ per second is smaller than the electric current in units of $e$       
per second ($e$ is the electron charge). Such a situation naturally emerges
in partially polarized systems since each electron carries the charge $e$ and
its spin projection on the $z$-axis is $\pm\hbar/2$. 
A proposal how to obtain a spin current \emph{exceeding} the charge current in 
a quantum wire was published by Sharma and Chamon \cite{Sharma} who 
considered a Luttinger liquid in the presence of a time-dependent 
magnetic field in a region of the size of an electron wavelength. 
In a very different physical context, a spin current without charge current was 
predicted for edge modes in the quantum Hall effect in graphene \cite{Abanin06}.
Pure spin currents can also flow in open circuits which cannot support charge currents
\cite{Pustilnik}.
In this paper, we show that the generation of a dc spin current exceeding the charge current 
is also possible in closed circuits without time-dependent magnetic fields.
The spin current can be generated in a spatially asymmetric 
Luttinger liquid system in the presence of an ac bias.
Interestingly, in a certain interval of low voltages the dc spin current 
grows as a negative power of the ac voltage when the voltage decreases.   

The paper is organized as follows: The next section contains a qualitative 
discussion. We briefly address the simplest case of non-interacting electrons, 
discuss its differences from the most interesting case of strong electron 
interaction and estimate at what conditions the effect can be observed. 
Section \ref{sec:bosonization} contains the details of the bosonization 
procedure which we use to treat the electron-electron interaction. 
In Section \ref{sec:rectification}  we calculate analytically the spin and 
charge rectification currents in the presence of a weak asymmetric potential. 
Numerical results for a simple model with strong asymmetric potential are 
discussed in Appendix \ref{sec:high_barrier}. 
Appendices \ref{sec:higher_orders} and \ref{sec:3rd_order} contain technical 
details of the perturbation theory employed in Sec. \ref{sec:rectification}.


\section{Model and Physics of the problem}
\label{sec:model}

\begin{figure}
\begin{center}
\includegraphics[width=\columnwidth]{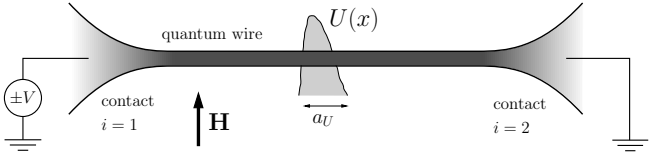}
\caption{
Sketch of the one-dimensional conductor connected to two electrodes
on both ends.
Currents are driven through a voltage bias $V$ that is applied on 
the left electrode while the right electrode is kept on ground.
The system is magnetized by the field $\mathbf{H}$. Electrons are 
backscattered off the asymmetric potential $U(x)$. $U(x)\ne 0$ in 
the region of size $a_U\sim 1/k_F$.
\label{fig:system}}
\end{center}
\end{figure}

The rectifying quantum wire is sketched in Fig.~\ref{fig:system}. 
It consists of a one-dimensional conductor with
a scatterer in the center of the system at $x=0$.
The scatterer creates an asymmetric potential $U(x)\ne U(-x)$.
The size of the scatterer $a_U\sim 1/k_F$ is of the order of the electron 
wavelength. A spin current can be generated only if time-reversal symmetry 
is broken. Thus, we assume that the system is placed in a uniform magnetic 
field $\mathbf{H}$.
The field defines the $S_z$ direction of the electron spins.
If the wire is sufficiently narrow then the effect of the magnetic field on the 
kinetic energy of electrons can be neglected and the field enters the problem 
only via its interaction with the spins.
At its two ends, the wire is connected to nonmagnetic electrodes, labeled 
by $i=1,2$. The left electrode, $i=1$, is controlled by an ac voltage source,
while the right electrode, $i=2$, is kept on ground. 

The magnetic field $\mathbf{H}$ breaks the symmetry between the two 
orientations of the electron spin.
In a uniform wire this would not result in a net spin current since the 
conductances of the spin-up and -down channels would be the same,\cite{CleanCurrent} 
$e^2/h$, and the spin currents of the spin-up and
-down electrons would be opposite. In the presence of a potential barrier 
such a cancellation does not occur \cite{Hikihara05}. In a system with strong 
electron interaction, the spatial asymmetry of the wire 
leads to an asymmetric $I-V$ curve,\cite{Feldman05} $I(V)\ne I(-V)$. 
Thus, an ac voltage bias generates spin and charge dc currents, $I^r_s$ and 
$I^r_c$. 

As we will see, the problem is most interesting in the case of strong electron 
interaction.
Before addressing that more difficult case, let us discuss what happens in the 
absence of electron interaction.
By non-interacting system we mean a wire in which electron-electron interaction 
is completely screened by the gates.
In such situation the charge density in the wire is not fixed but depends on 
the gate potential and the electrochemical
potentials of the leads. The leads define the chemical potentials $\mu_L$ and 
$\mu_R$ of the left- and right-moving electrons which are injected from the 
right and left reservoirs respectively. In what follows we will assume that 
the chemical potential $\mu_L=0$ and $\mu_R$ oscillates between $+eV$ and $-eV$. 
Thus, in the presence of the magnetic field $\mathbf{H}$, the Fermi energies 
counted from the band bottom equal $E_{F}^L(S_z)=E_F+2S_z\mu H$ for the 
left-moving electrons and $E_F^R(S_z)=E_F+\mu_R+2S_z\mu H$ for the right-movers, 
where $S_z=\pm 1/2$ is the electron spin projection, $E_F$ the Fermi level in 
the absence of the magnetic field and voltage bias and $\mu$ the electron 
magnetic moment. 

\begin{figure}
\begin{center}
\includegraphics[width=\columnwidth]{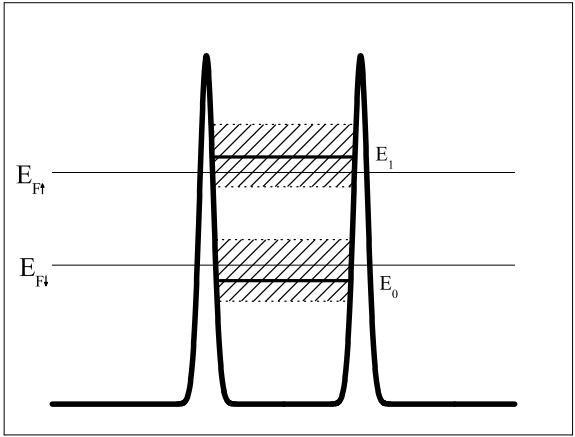}
\caption{
Double-well potential with quasistationary levels. 
The transmission coefficient is maximal in the shaded regions.
The narrow potentials $u_1(x)$ and $u_2(x)$ are centered
at the positions $x=0$ and $x=a$ ($a < k_F^{-1}$), respectively,
and are modeled by $\delta$-functions in Eq.~\eqref{a1_add}.
\label{fig:LR}}
\end{center}
\end{figure}

As we will see, in a strongly interacting system the form of the potential 
barrier $U(x)$ plays little role. However, it is crucial in the non-interacting 
case. Let us choose $U(x)$ in the form of the double potential barrier so that 
quasistationary levels $E_n$, $n=0,1,\ldots$ are present (Fig.~\ref{fig:LR}). 
Thus, in the non-interacting case we consider a resonant tunneling 
diode\cite{resonant}. The spin and charge currents as the functions of the 
chemical potential $\mu_R$ are
\begin{equation}
	\label{I_c_add}
	I_c(\mu_R)=I(\mu_R,1/2)+I(\mu_R,-1/2); 
\end{equation}
\begin{equation}
	\label{I_s_add}
	I_s(\mu_R)=\frac{\hbar}{2e}[I(\mu_R,1/2)-I(\mu_R,-1/2)], 
\end{equation}
where $e$ is the electron charge,
\begin{equation}
	\label{3-0_add}
	I(\mu_R,S_z)=\frac{e}{h}\int_{E_F^L(S_z)}^{E_F^R(S_z)}dE T(E);
\end{equation}
and $T(E)$ is the transmission coefficient. 
The transmission coefficient is small far from the energies of the 
quasistationary levels $E=E_n$ and increases as $E$ approaches $E_n$.
Obviously, if one applies a dc voltage $eV=\mu_R$ then the charge current, 
expressed in units of $e$ per second, exceeds the spin current in units of 
$\hbar/2$ per second. The situation changes in the presence of an ac bias. 
The dc currents generated by an ac voltage bias can be estimated
as $I_{c/s}^r=[I_{c/s}(V)+I_{c/s}(-V)]/2$. Let us now assume that the 
magnetic field is tuned such that the Fermi level of the spin-down electrons
$E_F-\mu H$ is close to the quasistationary level $E_0$ and exceeds $E_0$ 
while the Fermi level of the spin-up electrons $E_F+\mu H$ is close to $E_1$ 
and lies below $E_1$. Let also $eV$ be smaller than the distances, 
$|E_0-E_F+\mu H|$ and $|E_1-E_F-\mu H|$, between the Fermi levels and 
quasistationary levels in the absence of the voltage bias.
In addition, we assume that $T(E_F+\mu H)=T(E_F-\mu H)$.
From the energy dependence of the transmission coefficient $T(E)$ near the 
resonant levels, one finds that $I(eV,1/2)>I(eV,-1/2)$ and 
$|I(-eV,1/2)|<|I(-eV,-1/2)|$. Hence, $I^r_s>\frac{\hbar}{2e}I_c^r$. 
By an appropriate choice of parameters, one can produce any ratio of 
the spin and charge rectification currents.

Note that the rectification effect for non-interacting electrons is 
possible even if the potential $U(x)$ is symmetric.
The asymmetry of the system, necessary for rectification, is introduced by 
the applied voltage bias. The charge density injected into the wire
from the leads is proportional to $\mu_L+\mu_R$ and hence is different 
for the opposite signs of the voltage. If the injected charge density were 
independent of the voltage sign, i.e. $\mu_R$ oscillated between $eV/2$ and 
$-eV/2$ and $\mu_L=-\mu_R$ oscillated between $-eV/2$ and $eV/2$, then the 
rectification effect would be impossible for non-interacting electrons.
This follows from Eqs. (\ref{I_c_add}-\ref{3-0_add}) and the fact that 
the transmission coefficient $T(E)$ is independent of the direction of 
the incoming wave for non-interacting particles\cite{Landau}. In the 
presence of electron repulsion both the asymmetry of the potential and 
the voltage dependence of the injected charge contribute to the rectification 
current. It turns out, that in the case of strong electron interaction, the 
rectification effect due to the asymmetry of the potential barrier dominates.

The above example is based on a special form of the potential barrier 
in the wire and assumes that the magnetic field and chemical potentials 
are tuned in order to obtain the desired effect.  As shown below, in the 
presence of strong repulsive electron interaction no tuning is necessary 
and no quasistationary states are needed to obtain the spin current
which is greater than the charge current. In fact, the spin rectification 
effect is possible even for weak asymmetric potentials $U(x)$. This can 
be understood from the following toy model (a related model for rectification 
in a two-dimensional electron gas was studied in Ref.~\onlinecite{Scheid}): 
Let there be no uniform magnetic 
field $\mathbf{H}$ and no asymmetric potential $U(x)$. Instead, both 
right$\leftrightarrow$left and spin-up$\leftrightarrow$spin-down symmetries 
are broken by a weak coordinate-dependent magnetic field $B_z(x)\ne B_z(-x)$, 
which is localized in a small region of size $\sim 1/k_F$ (we do not include 
the components $B_{x,y}$ in the toy model). Let us also assume that the 
spin-up and -down electrons do not interact with the electrons of the opposite
spin. Then the system can be described as the combination of two 
spin-polarized one-channel wires with opposite spin-dependent potentials 
$\pm\mu B_z(x)$, where $\mu$ is the electron magnetic moment. According to 
Ref. \onlinecite{Feldman05} an ac bias generates a rectification current in each 
of those two systems and the currents are proportional to the cubes of the 
potentials $(\pm \mu B_z)^3$.
Thus, $I^r_\uparrow=-I^r_\downarrow$. Hence, no net charge current 
$I^r=I^r_\uparrow+I^r_{\downarrow}$ is generated in the leading order. 
At the same time, there is a nonzero spin current in the third order in $B_z$.
A similar effect is present in a more realistic Luttinger liquid model 
considered below.

\begin{figure}
\begin{center}
\includegraphics[width=\columnwidth]{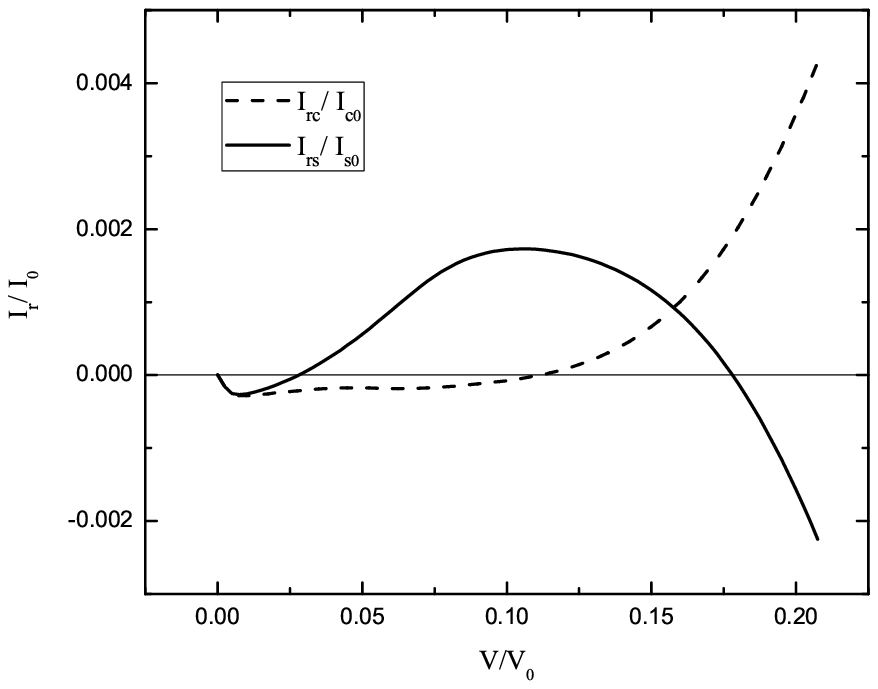}
\caption{Normalized charge rectification
current $I_{c}/I_{c0}$ and spin rectification current $I_{s}/I_{s0}$
versus applied voltage $V/V_0$ for interacting electrons with
$E_F=100\epsilon_0$, $\mu H=25\epsilon_0$,
$\gamma=12.6\epsilon_0 a/e$, $u_{1}=25\epsilon_0a$ and
$u_{2}=50\epsilon_0a$, where $\epsilon_0=\hbar^2/ma^2$
(see Fig.~\ref{fig:LR} and Appendix~\ref{sec:high_barrier}).
$I_{c0}=50e\epsilon_{0}/\hbar$, $I_{s0}=25\epsilon_{0}$, and
$V_{0}=50\epsilon_{0}/e$ are arbitrary reference currents
and voltage.
\label{fig:interact}}
\end{center}
\end{figure}

The main focus of this paper is on the case of weak asymmetric potentials. A
simple model with strong impurities is studied in Appendix \ref{sec:high_barrier}.
In Figs.~\ref{fig:noninteract} and \ref{fig:interact} we have represented
the results from a numerical evaluation for the spin and charge currents
$I_{s,c}^r$ for the potential shown in Fig.~\ref{fig:LR}.
Fig.~\ref{fig:noninteract}
shows the non-interacting case. In Fig.~\ref{fig:interact} we represent the
case of strong 
electron interaction. We have chosen parameters (explained in the figure
captions) such that $I^r_c$ is smaller than $I^r_s$ for a range of the
applied voltage. Further information on the numerical approach is given
in Appendix~\ref{sec:high_barrier}.

Transport in a strongly interacting system in the presence of a strong
asymmetric potential $U(x)$ is a difficult problem which cannot be solved
analytically and is sensitive to a particular choice of the potential. 
As we have mentioned, Appendix \ref{sec:high_barrier} contains the numerical 
analysis of a simple model of interacting electrons with a strong potential 
barrier. On the other hand, the interacting problem can be solved analytically 
in the limit of a weak potential $U(x)$ with the help of the bosonization and 
Keldysh techniques (Sec.~\ref{sec:bosonization}). We will see that the rectification current 
exhibits a number of universal features, independent of the form of the 
potential $U(x)$. In particular, in a wide interval of interaction strength, 
the spin rectification current can exceed the charge rectification current 
for an arbitrary shape of the asymmetric potential barrier.

Rectification is a nonlinear transport phenomenon. Thus, it cannot be observed 
at low voltages at which the $I-V$ curve is linear and hence symmetric. 
In Luttinger liquids the $I-V$ curve is nonlinear at $eV>k_B T$,
where $T$ is the temperature \cite{Kane92}. We will concentrate on the limit 
of the zero temperature which corresponds to the strongest rectification. 
We expect qualitatively the same behavior at $T\sim V$. 
At higher temperatures the charge and spin rectification effects disappear.
Since the temperatures of the order of millikelvins can be achieved with 
dilution refrigeration, the rectification effect is possible even for the 
voltages as low as $V\lesssim 1~\mu$V.

In this paper we focus on the low-frequency ac bias. We define the 
rectification current as the dc response to a low-frequency square voltage 
wave of amplitude $V$: 
\begin{equation}
	\label{5_add}
	I_s^r(V)=[I_s(V)+I_s(-V)]/2,
\end{equation}
\begin{equation}
	\label{6_add}
	I_c^r(V)=[I_c(V)+I_c(-V)]/2.
\end{equation}
The above dc currents express via the currents of spin-up and -down electrons:
$I_c^r=I_\uparrow^r+I_\downarrow^r$, 
$I_s^r=(\hbar/2e)[I_\uparrow^r-I_\downarrow^r]$. 
The spin current exceeds the charge current if the signs of  $I^r_{\uparrow}$ 
and $I^r_{\downarrow}$ are opposite. Equations (\ref{5_add},\ref{6_add}) 
for the dc-currents do not contain the frequency $\omega$ of the ac-bias. 
They are valid as long as the frequency 
\begin{equation}
	\label{7_add}
	\omega<eV/\hbar.
\end{equation} Indeed, as shown below, 
the rectification current is determined by electron backscattering
off the asymmetric potential. Hence, one can neglect the time-dependence of 
the ac voltage in Eqs. (\ref{5_add},\ref{6_add}) if the period of the ac bias 
exceeds the duration $\tau$ of one backscattering event.
The time $\tau\sim\tau_{\rm travel}+\tau_{\rm uncertainty}$ includes two 
contributions. $\tau_{\rm travel}$ is the time of the electron travel across 
the potential barrier. $\tau_{\rm uncertainty}$ comes from the uncertainty 
of the energy of the backscattered particle. If the barrier amplitude 
$U(x)<E_F$ and the barrier occupies a region of size 
$a_U\sim 1/k_F$ then $\tau_{\rm travel}\sim 1/k_Fv\sim \hbar/E_F$, where 
$v\sim \hbar k_F/m$ is the electron velocity. The energy uncertainty $\sim eV$ 
translates into $\tau_{\rm uncertainty}\sim\hbar/eV$. Thus, for $eV<E_F$ one 
obtains the condition (\ref{7_add}). The same condition 
can be derived with the approach of Appendix A of 
Ref. \onlinecite{voltage-keldysh}
and emerges in a related problem \cite{Feldman03}. Note that for realistic 
voltages the low-frequency condition (\ref{7_add}) allows rather high 
frequencies. Even for $V\sim 1~\mu$V the maximal $\omega\sim 1$ GHz.

There remains the question of the asymmetric impurity: We require a
potential $U(x)$ that is localized within $\sim 1/k_F$. 
A possible realization is to generate two different (symmetric) local 
potentials by two gates within a distance $\sim1/k_F$ or an electric potential 
created by an asymmetric gate of size $\sim 1/k_F$ placed at the distance 
$\sim 1/k_F$ from the wire. 
Electron densities of $\rho \sim 10^{11} \mathrm{cm}^{-2}$ are possible 
nowadays in 2-dimensional electron gases, yielding $1/k_F$ up to several 10 nm.
Confinement in a one-dimensional wire will reduce the electron density further
so that this number may increase further.
Modern techniques allow placing electric gates of widths of $\sim 20$ nm
at distances of $\sim 20 - 50$ nm. A realization of an asymmetric potential
in this way is, therefore, within the reach. 
Alternatively, in the case of shorter electron wave-length, it should be 
possible to place an asymmetrically shaped STM tip close to the wire. 
An applied bias would yield an asymmetric scattering potential. With such a 
tip the asymmetry cannot be directly tuned, but most of our predictions are 
not sensitive to the precise shape of the potential. Certainly, an asymmetric 
potential may simply emerge by chance due to the presence of two point 
impurities of unequal strength at the distance $\sim 1/k_F$.


\section{Bosonization and Keldysh Technique}
\label{sec:bosonization}

At $\omega<V$, the calculation of the rectification currents reduces to the 
calculation of the stationary contributions to the dc $I-V$ curves $I_s(V)$ 
and $I_c(V)$ that are even in the voltage $V$.
We assume that the Coulomb interaction between distant charges is screened 
by the gates. This will allow us to use the standard Tomonaga-Luttinger model 
with short range interactions. \cite{Kane92} Electric fields of external 
charges are also assumed to be screened.
Thus, the applied voltage reveals itself only as the difference
of the electrochemical potentials $E_1$ and $E_2$ of the
particles injected from the left and right reservoirs.
We assume that one lead is connected to the ground so
that its electrochemical potential $E_2=E_F$ is fixed. The
electrochemical potential of the second lead $E_1=E_F+eV$ is controlled by 
the voltage source (see Fig.~\ref{fig:system}).
Since the Tomonaga-Luttinger model captures only low-energy physics, we 
assume that $eV< E_F$, where $E_F$ is of the order of the bandwidth. 
Rectification occurs due to backscattering off the asymmetric potential 
$U(x)$. We will assume that the asymmetric potential is weak, $U(x)< E_F$.
This will enable us to use perturbation theory. 

We assume that the magnetic field $\mathbf{H}$ couples 
only to the electron spin and we neglect the correction $-e \mathbf{A}/c$ to 
the momentum in the electron kinetic energy. Indeed, for a uniform field
one can choose $\mathbf{A}\sim y$, where the $y$-axis is orthogonal to the 
wire, and $y$ is small inside a narrow wire.
As shown in Ref.~\onlinecite{Hikihara05}, such a system allows a formulation within
the bosonization language and, in the absence of the asymmetric potential, 
can be described by a quadratic bosonic Hamiltonian
\begin{equation} \label{eq:H0}
	H_0 = 
	\sum_{\nu,\nu' = L,R}\sum_{\sigma,\sigma'=\uparrow,\downarrow}
	\int dx 
	\bigl(\partial_x \phi_{\nu\sigma}\bigr) 
	\mathcal{H}_{\nu\sigma,\nu'\sigma'}
	\bigr(\partial_x \phi_{\nu'\sigma'}\bigr),
\end{equation}
where $\sigma$ is the spin projection and $\nu=R,L$ labels the left and right 
moving electrons, which are related to the boson fields $\phi_{\nu\sigma}$ as 
$\psi^\dagger_{\nu\sigma}(x) \sim \eta^\dagger_{\nu\sigma}
\mathrm{e}^{\pm i(k_{F\nu\sigma} x + \phi_{\nu\sigma}(x))}$
with $\pm$ for $\nu = R,L$.
The operators $\eta^\dagger_{\nu\sigma}$ are the Klein factors adding a 
particle of type $(\nu,\sigma)$ to the system, and $k_{F\nu\sigma}/\pi$ is 
the density of $(\nu,\sigma)$ particles in the system.
The densities of the spin-up and -down electrons are different since 
the system is polarized by the external magnetic field.
The $4 \times 4$ matrix $\mathcal{H}$ describes the electron-electron
interactions. In the absence of spin-orbit interactions,
$L \leftrightarrow R$ parity is conserved and we can 
introduce the quantities $\phi_\sigma = \phi_{L\sigma}+\phi_{R\sigma}$
and $\Pi_\sigma = \phi_{L\sigma}-\phi_{R\sigma}$ such that the Hamiltonian
decouples into two terms depending on $\phi_\sigma$ and $\Pi_\sigma$
only.
In the absence of the external field, this 
Hamiltonian would further be diagonalized by the combinations 
$\phi_{c,s} \propto \phi_{\uparrow} \pm \phi_{\downarrow}$, and similarly for
$\Pi_{c,s}$, expressing the spin and charge separation.
This is here no longer the case because of the external magnetic field. 
If we focus on the $\phi$ fields only
(as $\Pi$ will not appear in the operators describing backscattering off 
$U(x)$), the fields diagonalizing the Hamiltonian, $\tilde{\phi}_{c,s}$, have 
a more complicated linear relation to $\phi_{\uparrow,\downarrow}$, which 
we can write as
\begin{equation} \label{eq:transf}
	\begin{pmatrix} \phi_\uparrow \\ \phi_\downarrow \end{pmatrix}
	= 
	\begin{pmatrix} 
		\sqrt{g_c}[1+\alpha] &  \sqrt{g_s} [1+\beta] \\
		\sqrt{g_c}[1-\alpha] & -\sqrt{g_s} [1-\beta]
	\end{pmatrix}
	\begin{pmatrix} \tilde{\phi}_c \\ \tilde{\phi}_s \end{pmatrix},
\end{equation}
and which corresponds to the matrix $\hat{A}^T$ of Ref.~\onlinecite{Hikihara05}.
The normalization has been chosen such that the propagator of the 
$\tilde{\phi}$ fields with respect to the Hamiltonian \eqref{eq:H0} evaluates 
to 
$\langle \tilde{\phi}_{c,s}(t_1) \tilde{\phi}_{c,s}(t_2) \rangle 
= - 2 \ln(i(t_1-t_2)/\tau_c+\delta)$, where $\delta > 0$ is an infinitesimal quantity
and $\tau_c \sim \hbar /E_F$ the ultraviolet cutoff time.
For non-interacting electrons without a magnetic field,
$g_c=g_s = 1/2$. $g_c < 1/2$ ($> 1/2$) for repulsive (attractive) interactions. 
The interaction constants depend on microscopic details and the magnetic field.
The dimensionless parameter which controls the interaction strength is the ratio of 
the potential and kinetic energies
of the electrons. This ratio grows as the charge density decreases and hence
lower electron densities correspond to stronger repulsive interaction.
In the absence of the magnetic field, terms in \eqref{eq:H0} in the form of 
$\exp(\pm 2i\sqrt{g_s}\phi_s)$ may become relevant and open a spin gap
for $g_s < 1/2$. 
In our model they can be neglected since they are suppressed by the rapidly 
oscillating factors $\exp(\pm2i[k_{F\uparrow}-k_{F\downarrow}]x)$. 
It is convenient to model the leads as the regions near the right 
and left ends of the wire without electron interaction \cite{CleanCurrent}.

Backscattering off the impurity potential $U(x)$ is described by the following 
contribution to the Hamiltonian\cite{Kane92} $H = H_0 +H'$:
\begin{equation} \label{eq:backscattering}
	H' = \sum_{n_\uparrow,n_\downarrow} 
	U(n_\uparrow,n_\downarrow)
	\mathrm{e}^{i n_\uparrow \phi_{\uparrow}(0) + i n_\downarrow \phi_{\downarrow}(0)},
\end{equation}
where the fields are evaluated at the impurity position $x=0$ and 
$U(n_\uparrow,n_\downarrow)=U^* (-n_\uparrow,-n_\downarrow)$
since the Hamiltonian is Hermitian. The fields $\Pi$ do not enter the above 
equation due to the conservation  of the electric charge and the 
$z$-projection of the spin.
The Klein factors are not written because they drop out in the perturbative 
expansion.
$U(n_\uparrow,n_\downarrow)$ are the amplitudes of backscattering of 
$n_\uparrow$ spin-up and $n_\downarrow$ spin-down particles
with $n_\sigma > 0$ for $L \to R$ and $n_\sigma <0$ for $R\to L$ scattering. 
$U(n_\uparrow,n_\downarrow)$  can be estimated as\cite{Kane92}
$U(n_\uparrow,n_\downarrow) \sim k_F \int dx U(x) 
\mathrm{e}^{i n_\uparrow 2k_{F\uparrow}x+i n_\downarrow 2k_{F\downarrow}x}
\sim U$,
where $U$ is the maximum of $U(x)$. 
In the case of a symmetric potential, $U(x)=U(-x)$, the coefficients 
$U(n_\uparrow,n_\downarrow)$ are real.

The spin and charge current can be expressed as
\begin{equation}
	I_{s,c} = L_{s,c}^1 + R_{s,c}^1=L_{s,c}^2+R_{s,c}^2,
\end{equation} 
where $L_{s,c}^i$ and $R_{s,c}^i$ denote the current of the left-
and right-movers near electrode $i$, respectively (see Fig.~\ref{fig:LR}).
For a clean system ($U(x)=0$), the currents obey\cite{CleanCurrent}
$R_{s,c}^1=R_{s,c}^2$, 
$L_{s,c}^1=L_{s,c}^2$ and $I_c = 2e^2V/h$, $I_s=0$.
With backscattering off $U(x)$, particles are transferred between $L$ 
and $R$ in the wire, and hence $R_{s}^2=R_{s}^1+dS_R/dt$, 
$R_c^2=R_c^1+dQ_R/dt$, where $Q_R$ and $S_R$ denote the total charge 
and the $z$-projection of the spin of the right-moving electrons
\cite{Feldman03}.
The currents $L_{s,c}^2$ and $R_{s,c}^1$ are determined by 
the leads (i.e. the regions without electron interaction in our 
model \cite{CleanCurrent}) and remain the same as in the absence
of the asymmetric potential. Thus, the spin and charge current can 
be represented as $I_c=2e^2V/h+I_c^{bs}$ and $I_s=I_s^{bs}$, where 
the backscattering current operators are \cite{Feldman05,BB05b,Kane92}
\begin{align} 
\label{eq:I_c}
	\hat{I}_c^{bs} 
	&= {d\hat Q_R}/{dt}=i[H,\hat{Q}_R]/\hbar  \nonumber \\
	&=\frac{-ie}{\hbar}
	\sum_{n_\uparrow,n_\downarrow}(n_\uparrow+n_\downarrow)
	U(n_\uparrow,n_\downarrow)
	\mathrm{e}^{i n_\uparrow \phi_{\uparrow}(0) + i n_\downarrow \phi_{\downarrow}(0)},
\\
\label{eq:I_s}
	\hat{I}_s^{bs}  
	&={d\hat{S}_R}/{dt}
	\nonumber \\
	&= 
	-\frac{i}{2}\sum_{n_\uparrow,n_\downarrow}(n_\uparrow-n_\downarrow)
	U(n_\uparrow,n_\downarrow)
	\mathrm{e}^{i n_\uparrow \phi_{\uparrow}(0) + i n_\downarrow \phi_{\downarrow}(0)}. 
\end{align}
The calculation of the rectification currents reduces to the calculation of 
the currents \eqref{eq:I_c}, \eqref{eq:I_s} at two opposite values of the 
dc voltage.

To find the backscattered current we use the Keldysh technique
\cite{Keldysh}. We assume that at $t=-\infty$ there is no backscattering
in the Hamiltonian ($U(x)=0$), and then the backscattering is
gradually turned on. Thus, at $t=-\infty$, the numbers $N_L$ and $N_R$ of the 
left- and right-moving electrons conserve separately: 
The system can be described by a partition function with two chemical 
potentials $E_1=E_F+eV$ and $E_2=E_F$ conjugated with the
particle numbers $N_R$ and $N_L$. This initial state determines
the bare Keldysh Green functions. 

We will consider only the zero temperature limit. It is convenient
to switch \cite{Feldman03} to the interaction representation 
$H_0\rightarrow H_0-E_1 N_R-E_2 N_L$. 
This transformation induces a time dependence in the electron 
creation and annihilation operators.
As the result each exponent in Eq.~\eqref{eq:backscattering} 
is multiplied by $\exp(ieVt[n_\uparrow+n_\downarrow]/\hbar)$.

In the Keldysh formulation \cite{Keldysh} the backscattering currents 
\eqref{eq:I_c}, \eqref{eq:I_s} are evaluated as
\begin{equation} \label{eq:I_Keldysh}
	I_{c,s}^{bs}
	= \langle 0 | \mathcal{S}(-\infty,0) 
	  \hat{I}_{c,s}^{bs} \mathcal{S}(0,-\infty) | 0 \rangle,
\end{equation}
where $|0\rangle$ is the ground state for the Hamiltonian $H_0$, 
Eq. \eqref{eq:H0}, and $\mathcal{S}(t,t')$ the evolution operator for 
$H'$ from $t'$ to $t$ in the interaction representation with respect 
to $H_0$. The result of this calculation depends on the elements of 
the matrix \eqref{eq:transf}, which describe the low-energy degrees of 
freedom and depend on the microscopic details. Several regimes
are possible \cite{BB05b} at different values of the parameters 
$g_s>0$, $g_c>0$, $\alpha$ and $\beta$. In this paper
we focus on one particular regime, in which the main contribution to 
the rectification current comes from backscattering operators 
$U(1,0)$, $U(0,-1)$ and $U(-1,1)$.


\section{Rectification currents}
\label{sec:rectification}

In order to calculate the current (\ref{eq:I_Keldysh}) we will expand 
the evolution operator in powers of $U(n,m)$.
Such perturbative approach is valid only if $U<E_F$. The details of 
the perturbative calculation are discussed in Appendix \ref{sec:3rd_order}.

The currents \eqref{eq:I_Keldysh} can be estimated using a 
renormalization group procedure \cite{Kane92}. 
As we change the energy scale $E$, the backscattering amplitudes 
$U(n_\uparrow,n_\downarrow)$ scale as 
\begin{equation}
	\label{scaling_add}
	U(n,m;E)\sim U(n,m)(E/E_F)^{z(n,m)},
\end{equation} 
where the scaling dimensions are
\begin{eqnarray} \label{eq:z}
	z(n,m) = n^2 [ g_c (1+\alpha)^2 + g_s (1+\beta)^2 ]
	       + m^2 [ g_c (1-\alpha)^2 & & \nonumber\\
	              +g_s (1-\beta)^2 ]
		   + 2 nm [ g_c (1-\alpha^2) - g_s (1-\beta^2) ] - 1 & & \nonumber\\
= n^2 A + m^2 B +2nm C-1.~~ & &
\end{eqnarray}
The renormalization group (RG) stops at the scale of the order $E\sim eV$. 
At this scale the backscattering current can be represented as 
$I_{c,s}^{bs} = V r_{c,s}(V)$, where the effective reflection coefficient 
$r_{c,s}(V)$ is given by the sum of contributions 
of the form\cite{Feldman05,Kane92}
\begin{equation}
	\label{orders_add}
	(\mathrm{const})U(n_1,m_1;E=eV)U(n_2,m_2;eV)\dots U(n_p,m_p;eV).
\end{equation}
Such a perturbative expansion can be used as long as 
\begin{equation}
	\label{condition_aa}
	U(n,m;E=eV)<E_F
\end{equation}
for every $(n,m)$. This condition defines the RG cutoff voltage $V^*$ such that
\begin{equation}
	\label{V_add}
	U(n_0,m_0;E=eV^*)=E_F
\end{equation}
for the most relevant operator $U(n_0,m_0)$. The RG procedure cannot be 
continued to lower energy scales $E<V^*$.

One expects that the leading contribution to the backscattering current 
emerges in the second order in $U(n,m)$, if the above condition is satisfied. 
The leading contribution to the {\it rectification current}
may however emerge in the third order.
Indeed, the second order contributions to the charge current were computed 
in Ref. \onlinecite{Kane92}. The spin current can be found
in exactly the same way. The result is 
\begin{equation} \label{eq:2order}
	I_{c,s}^{bs(2)}(V)
	\sim
	\sum (\mathrm{const}) 
	|U(n,m)|^2|V|^{2z(n,m)+1}\mathrm{sign}(V).
\end{equation}
If the (unrenormalized) $U(n,m)$ were independent of the voltage, the above 
current would be an odd function of the bias and hence would not contribute 
to the rectification current. The backscattering amplitudes depend\cite{Kane92} 
on the charge densities $k_{F\nu\sigma}$ though, which in turn depend on the 
voltage in our model \cite{Feldman05}. The voltage-dependent corrections to 
the amplitudes are linear in the voltage at low bias $eV\ll E_F$. Hence, the 
second order contributions to the rectification currents scale as 
$U^2|V|^{2z(n,m)+2}$ (see Appendix~\ref{sec:higher_orders}). 
The additional factor of $V$ makes the second order contribution smaller than 
the leading third order contribution \eqref{eq:spin} at sufficiently high 
impurity strength $U\ll E_F$ (as shown in Appendix~\ref{sec:higher_orders}, 
$U/E_F$ must exceed $[V/E_F]^{1+z(1,0)-z(0,1)-z(1,-1)}$). 
Note that the second order contribution to the rectification current is 
nonzero even for a symmetric potential $U(x)$ and emerges solely due to the 
voltage dependence of the injected charge density (cf. Sec.~\ref{sec:model}). 
The leading third order contribution emerges solely due to the asymmetry of 
the scatterer.

The main third order contribution comes from the three backscattering 
operators, most relevant in the renormalization group sense (small $z(n,m)$, 
Eq. (\ref{eq:z})).
They are identified in Appendix~\ref{sec:higher_orders}. Under conditions 
(\ref{1_add},\ref{0_add},\ref{2_add},\ref{-1_add},\ref{4b_add},\ref{5b_add}),
the most relevant operator is $U(1,0)$, the second most relevant 
$U(0,-1)$, and the third most relevant $U(-1,1)$. The cutoff voltage $V^*$ 
is determined by the scaling dimension $z(1,0)$, $eV^*\sim E_F(U/E_F)^{1/[1-A]}$.
The leading non-zero 
third order contributions to the spin and charge currents come from 
the product of the above three operators in the Keldysh perturbation 
theory (see Appendix B). This leads to
\begin{equation} \label{eq:scaling}
	I^{bs}_{c,s}\sim U^3 V^{2(A+B-C-1)}.
\end{equation}
This contribution dominates the spin rectification  current at 
\begin{equation}
	\label{domination_add}
	E_F(U/E_F)^{1/[2+2C-2B]}\equiv eV^{**}>eV>eV^*
\end{equation}
as is clear from the comparison with the leading
second order contribution $I^{bs}_{s,2}\sim U^2V^{2A}$, Eq. (\ref{I_2_add}).
Interestingly, the current \eqref{eq:scaling} grows as the voltage decreases 
in the regime
(\ref{1_add},\ref{0_add},\ref{2_add},\ref{-1_add},\ref{4b_add},\ref{5b_add}). 

However, does the current \eqref{eq:scaling} actually contribute to the 
rectification effect? 
In general, \eqref{eq:scaling} is the sum of odd and even functions of the 
voltage and only the even part is important for us. One might naively expect 
that such a contribution has the same order of magnitude for the spin and 
charge currents. A direct calculation shows, however, that this is not the 
case and the spin rectification current is much greater than the charge 
rectification current.

In order to calculate the prefactors in the right hand side of 
Eq. \eqref{eq:scaling} one has to employ the Keldysh formalism.
The details are explained in Appendix~\ref{sec:3rd_order}.
Here let us shortly summarize the essential steps:
The third order Keldysh contribution reduces to the integral of 
$P(t_1,t_2,t_3)=\langle T_c\exp(i\phi_\uparrow(t_1)+ieVt_1/\hbar)
\exp(-i\phi_\downarrow(t_2)-ieVt_2/\hbar)
\exp(i[-\phi_\uparrow(t_3)+\phi_\downarrow(t_3)])\rangle$
over $(t_1-t_3)$ and $(t_2-t_3)$, where $T_c$ denotes time ordering along 
the Keldysh contour $-\infty \to 0 \to -\infty$ and the angular 
brackets denote the average with 
respect to the ground state of the non-interacting Hamiltonian \eqref{eq:H0}. 
The integration can be performed analytically as discussed in 
Appendix~\ref{sec:3rd_order}.
One finds
\begin{align} 
\label{eq:charge}
	I^{bs}_c
	= &\frac{16e\tau_c^2}{\pi\hbar^3} \mathrm{sign}(eV)
	\left| \frac{eV\tau_c}{\hbar}\right|^{a+b+c-2} 
	\Gamma(1-a)\Gamma(1-b)
	\nonumber \\
	&\times\Gamma(2-a-b-c) \Gamma(a+b-1)\sin\frac{\pi a}{2}
	\nonumber \\ &
	\times\sin\frac{\pi b}{2}\sin\frac{\pi(a+b)}{2}\sin\pi(a+b+c) 
	\nonumber\\ &
	\times\mathrm{Re}[U(1,0)U(-1,1)U(0,-1)],
\\
\label{eq:spin}
	I^{bs}_s
	= &\frac{16\tau_c^2}{\pi\hbar^2}\sin\frac{\pi a}{2}\sin\frac{\pi b}{2}
	\left|\frac{eV\tau_c}{\hbar}\right|^{a+b+c-2} 
	\nonumber \\
	&\times\Gamma(a+b-1)\Gamma(2-a-b-c)\Gamma(1-a)\Gamma(1-b)
	\nonumber \\
	&\times
	\Bigl\{\mathrm{Im}[U(1,0)U(-1,1)U(0,-1)]
		\cos\frac{\pi (a+b+c)}{2}
	\nonumber\\
		\times&[\sin\frac{\pi c}{2}+\cos\frac{\pi(a-b)}{2}
		\sin\frac{\pi(a+b+c)}{2}+\sin\frac{\pi(a+b)}{2}
	\nonumber \\ 
		\times&\cos\frac{\pi(a+b+c)}{2}]
		+\frac{1}{2}\mathrm{Re}[U(1,0)U(-1,1)U(0,-1)]
	\nonumber\\
		&\times\sin\frac{\pi(a-b)}{2}
		\sin{\pi(a+b+c)}\mathrm{sign}(eV)
	\Bigr\},
\end{align}
where $a=2A-2C$, $b=2B-2C$, $c=2C$ and $\tau_c\sim \hbar/E_F$ is the 
ultraviolet cutoff time. 
The charge current \eqref{eq:charge} is an \emph{odd} function of the voltage 
and hence does \emph{not} contribute to the rectification effect. The spin 
current \eqref{eq:spin} is a sum of an \emph{even} and odd functions and 
hence determines the spin rectification current
\begin{align} 
\label{eq:spin_add}
	I^{r}_s
	= &\frac{16\tau_c^2}{\pi\hbar^2}
	\sin\frac{\pi a}{2}\sin\frac{\pi b}{2}\cos\frac{\pi (a+b+c)}{2}
	\left|\frac{eV\tau_c}{\hbar}\right|^{a+b+c-2} 
  \nonumber \\
	&\times\Gamma(a+b-1)\Gamma(2-a-b-c)\Gamma(1-a)\Gamma(1-b)
  \nonumber \\
	&\times\mathrm{Im}[U(1,0)U(-1,1)U(0,-1)]
	[\cos\frac{\pi(a-b)}{2} \sin\frac{\pi(a+b+c)}{2}
  \nonumber \\ 
	&+\sin\frac{\pi c}{2}+\sin\frac{\pi(a+b)}{2}\cos\frac{\pi(a+b+c)}{2}].
\end{align}
It is non-zero if 
$\mathrm{Im}[U(1,0)U(-1,1)U(0,-1)]\ne 0$,
which is satisfied for asymmetric potentials.
The leading contribution to the charge rectification currents comes from other 
terms in the perturbation expansion.
Thus, we expect that in the region 
(\ref{1_add},\ref{0_add},\ref{2_add},\ref{-1_add},\ref{4b_add},\ref{5b_add}), 
the spin rectification 
current exceeds the charge rectification current
in an appropriate interval of voltages (\ref{domination_add}).
The difference between the spin and charge rectification current can be easily 
understood from the limit $A=B$.
In that case the charge current changes its sign under the transformation 
$U(1,0)\leftrightarrow U(0,-1)$, 
$V\rightarrow -V$. Since $U(1,0)$ and $U(0,-1)$ enter the current only in the 
combination $U(1,0)U(0,-1)$, 
this means that the charge current must be an odd function of the voltage bias. 
A similar argument shows that at $A=B$ the spin rectification current is an 
even function of the voltage in agreement with Eq. (\ref{eq:spin}).

\begin{figure}
\begin{center}
\includegraphics[width=\columnwidth]{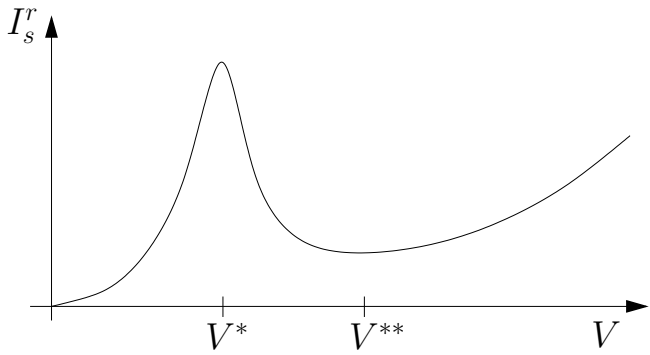}
\caption{
Qualitative representation of the spin rectification current.
The spin current exceeds the charge current and follows a power-law
dependence on the voltage with a negative exponent
in the interval of voltages $V^{*}<V<V^{**}$.
\label{fig:currents}}
\end{center}
\end{figure}

The voltage dependence of the spin rectification current is illustrated 
in Fig. \ref{fig:currents}. The expression \eqref{eq:scaling}
describes the current in the voltage interval $V^{**}>V>V^{*}$. 
In this interval the current \emph{increases} as the voltage decreases
in the regime 
(\ref{1_add},\ref{0_add},\ref{2_add},\ref{-1_add},\ref{4b_add},\ref{5b_add}). 
At lower voltages the perturbation theory breaks down. The current must 
decrease as the voltage decreases below $V^{*}$ and eventually reach 0 
at $V=0$. At higher voltages, $E_F> eV> eV^{**}$, the second order 
rectification current \eqref{eq:2order} dominates. The leading second 
order contribution $I^r_{s}\sim |U(1,0)|^2V^{2z(1,0)+2}$ grows as the 
voltage increases. The charge rectification current has the same order 
of magnitude as the spin current.

The Tomonaga-Luttinger model cannot be used for the highest voltage 
region $E_F\sim eV$.

It is easier to detect charge currents than spin currents. However, the 
measurement of the spin current can be reduced to the measurement of charge 
currents: Let us split the right end of the wire into two branches and place 
them in opposite strong magnetic fields so that only electrons with one 
spin orientation can propagate in each branch. 
If both branches are grounded, they still inject exactly the same charge and 
spin currents into the wire as one unpolarized
lead. However, the current generated in the wire will split between two 
branches into the currents of spin-up and spin-down electrons. If they are 
opposite then pure spin current is generated.


\section{Conclusions}
\label{sec:conclusions}

In this paper, we have shown that rectification in quantum 
wires in a uniform magnetic field can lead to a spin current that largely 
exceeds the charge current. 
The paper focuses on the regime of low voltages and weak asymmetric potentials 
in which the perturbation theory provides quantitatively exact predictions. 
Qualitatively the same behavior is expected up to $eV, U\sim E_F$.
The spin rectification effect is solely due to the properties of the wire 
and does not require time-dependent magnetic fields or spin polarized 
injection as from magnetic electrodes. 
The currents are driven by the voltage source only. In an interval of low 
voltages the spin current grows as the voltage decreases. In contrast to some 
other situations, the $z$-component of the total spin conserves and hence the 
dc spin current is constant throughout the system.


\begin{acknowledgments}
We thank J. B. Marston and D. Zumb\"{u}hl for many helpful discussions.
This work was supported in part by the NSF under grant numbers DMR-0213818, 
DMR-0544116, and PHY99-07949, and by Salomon Research Award. 
D.E.F. acknowledges the hospitality of the Aspen Center for Physics, of the 
MPI Dresden and of the KITP Santa Barbara where this work was completed.
\end{acknowledgments}


\appendix


\section{High potential barrier}
\label{sec:high_barrier}

In this appendix we first briefly consider the model of non-interacting 
electrons, Sec. \ref{sec:model}, and then a simple 
Hartree-type model for strongly interacting electrons.

\subsection{Model without interaction}

We consider non-interacting electrons in the presence of the potential 
\begin{equation}
	\label{a1_add}
	U(x)=u_1\delta(x)+u_2\delta(x-a).
\end{equation}
The transmission coefficient can be found from elementary quantum mechanics,
\begin{equation}
	\label{a2_add}
	T(E)=\frac{1}{(1-2s_1s_2\sin^2 ka)^2+(s_1+s_2+s_1s_2\sin 2ka)^2},
\end{equation}
where $E={\hbar^2 k^2}/{2m}$ and $s_i=mu_i/k\hbar^2$. The spin and charge 
rectification currents can be computed from Eqs. (\ref{I_c_add},\ref{I_s_add}). 
Fig. \ref{fig:noninteract} shows their voltage dependence for a certain choice 
of $u_1$, $u_2$, the voltage bias $V$ and the magnetic field $H$.

\subsection{Model with interaction}

It is difficult to find a general analytic expression for the current in the 
regime when both the electron interaction and potential barrier are strong. 
If all characteristic energies, $U$, $eV$, $\hbar^2/[ma^2]$ and the typical 
potential 
energy of an electron $E_P$, are of the order of $E_F$ then one can estimate 
the spin and charge rectification currents with dimensional analysis: 
$I^r_c\sim e E_F/\hbar$, $I_s^r\sim E_F$.

To obtain a qualitative picture of the interaction effects in the case of a high 
potential barrier (\ref{a1_add}), we restrict our discussion to a simple model 
in the spirit of the zero-mode approximation \cite{zma}. We assume that 
electrons move in a self-consistent Hartree-type field. In our ansatz the 
self-consistent field takes three different constant values $V_L$, $V_M$ and 
$V_R$ on the left of the potential barrier, between two $\delta$-function 
scatterers and on the right of the potential barrier. In the spirit of the 
Luttinger liquid model, we assume that the constants $V_L$, $V_M$ and $V_R$
are proportional to the average charge density in the respective regions, 
e.g., $V_M=\frac{\gamma}{a}\int_0^a dx \rho(x)$, where $\gamma$ is the 
interaction constant. 

A result is shown in Fig. \ref{fig:interact}.
We see that the voltage dependence of the spin and charge rectification 
current exhibits a behavior similar to the non-interacting case. 


\section{Estimation of higher perturbative orders}
\label{sec:higher_orders}

In this appendix we compare contributions to the rectification currents from 
different orders of perturbation theory.
We focus on the regime when the third order contribution dominates.
The appendix contains 5 subsections and has the following structure: 
1) We introduce a parametrization for the scaling dimensions (\ref{eq:z}).
2) We discuss the operators most relevant in the RG sense.
3) We determine at what conditions the second order contribution to the 
rectification current dominates. Subsection 3 also contains a lemma which is 
important in subsection 4.
4) We determine at what conditions the third order contribution to the current 
dominates.
5) We estimate the voltages and currents at which the spin rectification 
current can exceed the charge rectification current in realistic systems.

As shown in Refs.~\onlinecite{Feldman05} and \onlinecite{BB05b}, there 
are two effects leading to rectification in Luttinger liquids, which are here 
very shortly summarized: 
The \emph{density-driven} and the \emph{asymmetry-driven} rectification effects.
The former appears at second order in $U$. It appears because the backscattering
potential depends on the particle densities in the system, which in turn are
modified by the external voltage bias. The leading order backscattering 
currents are of the form\cite{Kane92} $I^{bs}(V) \sim \mathrm{sign}(V) U^2 |V|^\alpha$
so that the rectification currents, $I^r = [I^{bs}(V)+I^{bs}(-V)]$ vanish.
Due to the density dependence, however, an expansion of $U$ to linear order
in $V$ cancels the $\mathrm{sign}(V)$, and we obtain a rectification 
current $I^r \sim U^2 |V|^{\alpha+1}$.

The \emph{asymmetry-driven} rectification effect appears at third order
in $U$. It is due solely to the spatial asymmetry of the potential $U(x)$:
Due to backscattering off $U$, screening charges accumulate close to the 
impurity. Those create an electrostatic nonequilibrium backscattering
potential $W(x)$ for incident particles, leading to an effective potential
$\bar{U}(x) = U(x) + W(x)$. The spatial distribution of charges follows 
from the shape of $U(x)$ and the applied voltage bias. An asymmetric $U(x)$ 
leads to different electrostatic potentials for positive or negative bias, 
and hence to rectification. If we expand the current, 
$I^{bs}\sim \bar{U}^2 \sim U^2 + U W + \dots$, the asymmetry appears first 
at order $U W$. Since the charge density in the vicinity of the impurity is 
modified by the modification of the particle current through backscattering,
$W$ itself is (self-consistently) related to the 
backscattering current as $W \sim I^{bs}$. Hence $W \sim U^2$, so that the
asymmetric rectification effect appears first at third order in $U$, 
$I^r \sim U W \sim U^3$.

The main result of this paper are expressions for the currents
that result from the perturbation theory at third order in the 
impurity potential $U$. In this appendix we show that the considered
contribution indeed dominates the second and other third order expressions in
the region defined by Eq. 
(\ref{1_add},\ref{0_add},\ref{2_add},\ref{-1_add},\ref{4b_add},\ref{5b_add}).
In addition, we give the 
proof that higher perturbative orders $N\ge 4$ cannot exceed
these values in the considered range of the system parameters
$g_c,g_s, \alpha$ and $\beta$. Unless we want to emphasize the
correct dimensions, we set $E_F = 1$, $e = 1$ and $\hbar=1$ in this appendix.
We assume that $U<E_F$ and $eV<E_F$. 

An important observation is the following: In Eq. (\ref{orders_add}), 
$\sum_i n_i=\sum_i m_i =0$. This follows from the fact that in the absence 
of backscattering the numbers of right- and left-movers with different spin 
orientations conserve.

\subsection{Parametrization of scaling dimensions}

According to Eq. (\ref{eq:z})
\begin{equation} \label{z_ap_add}
	z(n,m) 
	= n^2 A + m^2 B +2nm C-1, 
\end{equation}
where
\begin{eqnarray}
	A=[ g_c (1+\alpha)^2 + g_s (1+\beta)^2 ] & & \\
	B=[ g_c (1-\alpha)^2 + g_s (1-\beta)^2 ] & & \\
	C=[ g_c (1-\alpha^2) - g_s (1-\beta^2) ] & &
\end{eqnarray}
Since $g_c$ and $g_s$ are positive,  $A$ and $B$ are also positive. $C$ can 
have any sign. It satisfies the inequality
\begin{equation}
	\label{sqrt_add}
	|C|<\sqrt{AB}.
\end{equation}
Indeed, $AB-C^2=4g_cg_s(1-\alpha\beta)^2>0$. Any values of $A,B>0$  and 
$-\sqrt{AB}<C<\sqrt{AB}$ are possible. For example, one can set 
$\alpha=\beta=(\sqrt{A}-\sqrt{B})/(\sqrt{A}+\sqrt{B})$,
$g_c=(\sqrt{A}+\sqrt{B})^2[1+C/\sqrt{AB}]/8$, 
$g_s=(\sqrt{A}+\sqrt{B})^2[1-C/\sqrt{AB}]/8$.

\subsection{Most relevant operators}

Depending on the values of $A$, $B$ and $C$ many different possibilities for 
relative importance of different backscattering operators $U(m,n)$ exist. 
In the paper we focus on the situation when the most relevant operator is 
$U(1,0)$, the second most relevant operator is $U(0,-1)$ and the third most 
relevant operator is $U(-1,1)$ (certainly, the scaling dimensions of the 
operators $U(n,m)$ and $U(-n,-m)$ are always the same). We will also assume 
that the operator $U(1,0)$ is {\it relevant} in the RG sense, i.e., $z(1,0)<0$.
The analysis of the situation in which $U(0,-1)$ is the most relevant operator, 
$U(1,0)$ is the second most relevant and $U(-1,1)$ is the third most relevant 
follows exactly the same lines. Similarly, little changes if $U(1,1)$ is the 
third most relevant operator.

The scaling dimensions of the three aforementioned operators are $A-1$, $B-1$ 
and $A+B-2C-1$. 
The following inequality must be satisfied in order for these operators to be 
most relevant backscattering operators:
$A-1<B-1<A+B-2C-1<[{\rm all~ other~ scaling~ dimensions~}]$. Hence
\begin{equation}
	\label{1_add}
	B>A>2C.
\end{equation}
Since $z(1,0)<0$,
\begin{equation}
	\label{0_add}
	A<1.
\end{equation}
When are all other operators less relevant?
We must consider three classes of operators: 1) $U(1,1)$; 2) $U(n,0)$ and 
$U(0,n)$ with $|n|>1$; 3) all other operators.

1) Since $z(1,1)=A+B+2C-1$, one finds
\begin{equation}
	\label{2_add}
	C>0.
\end{equation}
2) $z(0,n)=Bn^2-1>z(n,0)=An^2-1\ge 4A-1>A+B-2C-1$. Thus,
\begin{equation}
	\label{3_add}
	3A+2C>B.
\end{equation}
3) $z(n,m)-(A+B-2C-1)=An^2+Bm^2+2Cnm-(A+B-2C)\ge An^2+Bm^2 - C(n^2+m^2)-A - B+2C=(A-C)(n^2-1)+(B-C)(m^2-1)>0$ 
since $B-C>A-C>0$ in accordance with Eq. (\ref{1_add}), $|n|,|m|\ge 1$ and 
either $|n|$ or $|m|$ exceeds 1. 
Thus, case 3) gives no new restriction on $A$, $B$ and $C$.

\subsection{Second order contribution to the current}

When is the second order contribution to the rectification current dominant?
Any operator $U(n,m)$ can be represented as $\tilde U(n,m)+V U_1(n,m)+\dots$, 
where $U_1\sim \tilde U/E_F$.
Any second order contribution to the current which contains $\tilde U$ only is 
an odd function of the voltage bias.
Indeed, any such contribution is proportional to 
$\tilde U(n,m)\tilde U^*(n,m)=\tilde U(n,m)\tilde U(-n,-m)$. 
The transformation
$\tilde U\leftrightarrow \tilde U^*$, $V\rightarrow -V$ changes the sign of the 
current. At the same time, the transformation $\tilde U\leftrightarrow \tilde U^*$
cannot change the second order current at all. Hence, it is odd in the voltage. 
The same argument applies to any perturbative contribution which contains only 
$\tilde U$,  if every operator $\tilde U(n,m)$ enters in the same power as 
$\tilde U^*(n,m)$. {\it In particular}, if only two operators $\tilde U(n,0)$ 
and $\tilde U(0,m)$ and their conjugate enter then  the resulting current 
contribution is odd.

Thus, all second order contributions to the rectification current must contain $U_1$. 
As is clear from Eq. (\ref{eq:2order}), the leading second order contribution 
is proportional to the square of the most relevant operator, $|U(1,0)|^2$.
It scales as 
\begin{equation}
	\label{I_2_add}
	I_2\sim VU^2V^{2z(1,0)+1}\sim U^2V^{2A}.
\end{equation}

In this subsection we discuss at what conditions this contribution dominates 
for all $V>V^*$ (see Eq. (\ref{V_add})).
Since $U(1,0)$ is the most relevant operator, its renormalized amplitude 
$U(1,0;E=V)$ exceeds the renormalized amplitude of all other operators on 
every energy scale. At the same time it remains lower than 1 (i.e. $E_F$) 
for $V>V^*$. 
This certainly means that the renormalized amplitudes are smaller than 1 for 
all other operators too. Hence, the product of any operators is smaller than 
the product of any two of them and that product cannot exceed $U^2(1,0;E)$.
This guarantees that the second order current (\ref{I_2_add}) exceeds any 
second or higher order contribution which contains any operator $VU_1(n,m)$.
Thus, we have to compare $I_2$ with higher order contributions to the 
rectification current which contain $\tilde U$ only. Every such contribution 
is at least third order and contains at least one operator less relevant 
than $U(0,-1)$ [if it contains $U(\pm1,0)$ and $U(0,\pm 1)$ only then it must 
contain $VU_1$ as discussed above]. Thus, any rectification current 
contribution with $\tilde U$ only cannot exceed $U^3 V^{2z(1,0)+z(1,-1)+1}$.
Comparison with (\ref{I_2_add}) at $V\sim V^*$ leads to the condition 
\begin{equation}
	\label{I_2_condition}
	B>2C+1.
\end{equation}

\subsection{Third order contribution to the current}

The most interesting question is different. When does the third order 
contribution dominate the rectification current? 
We will focus on the third order contribution $I_3$ proportional to 
$U(1,0)U(0,-1)U(-1,1)$ at $V\sim V^*$. Note that this contribution is 
proportional to $\sim V^{2(A+B-C-1)}$ and hence scales as a negative power 
of the voltage, if
\begin{equation}
	\label{-1_add}
	A+B<C+1.
\end{equation}
At $V\sim V^*$, $U\sim V^{1-A}$. Thus, 
\begin{equation}
	\label{I_3_add}
	I_3(V=V^*)\sim V^{2B-A-2C+1}.
\end{equation}
We need to compare $I_3$, Eq. (\ref{I_3_add}), with the following types of 
contributions:
1) those containing at least three different operators 
[we treat a pair of $U(n,m)$ and $U(-n,-m)=U^*(n,m)$ as one operator];
2) those containing only one type of operators;
3) those containing  two types of operators.

Cases 1) and 2) are easy.

1) $I_3$ contains the product of the three most relevant operators and hence 
always exceeds the product of any other three different operators at any 
energy scale $E_F>V>V^*$.  Any contribution with three different operators 
is the product of three different operators times perhaps some other 
combination of operators which cannot exceed 1 at $E_F>V>V^*$. 
Hence it is smaller than $I_3.$

2) Any contribution to the rectification current with only one type of 
operators must contain $VU_1$. As discussed in the previous subsection, 
the leading contribution of such type emerges in the second order. It is 
$I_2$, Eq. (\ref{I_2_add}). At $V\sim V^*$, $I_2(V=V^*)\sim V^2$.  
The condition $I_2(V^*)<I_3(V^*)$ means that 
\begin{equation}
	\label{4b_add}
	2B<A+2C+1.
\end{equation}

3) We have to consider three possibilities: 
3.1) one operator has the form $U(n,0)$ and the second operator has the form 
$U(k,m)$, $m\ne 0$ or one operator has the form $U(0,m)$ and the other one has 
the form $U(n,k)$, $n\ne 0$; 
3.2) both operators have the form $U(n_i,0)$ or both operators have the form 
$U(0,m_i)$; 
3.3) both operators have the form $U(n_i,m_i)$ with $n_i,m_i\ne 0$.

3.1). Let us assume that one operator has the form $U(n,0)$ and the second one 
is $U(k,m)$. The case of the operators $U(0,m)$ and $U(n,k)$ can be considered 
in exactly the same way. We must have the same number of operators $U(k,m)$ 
and $U(-k,-m)$ in the perturbative contribution since the sum of the second 
indexes $\pm m$ must be 0. [The other cases are covered in 3.3).]
From the analysis of the sum of the first indexes one concludes that the 
operators $U(n,0)$ and $U(-n,0)$ also enter in the same power. It follows 
from the previous subsection that the perturbative contribution must contain 
at least one $U_1$ operator and hence is smaller than $I_2$. Hence, it is also 
smaller than $I_3$.

3.2) We will focus on the case when both operators have the form $U(n_i,0)$. 
The case when both operators have the form $U(0,m_i)$ is very similar and 
does not lead to a new restriction on $A$, $B$ and $C$. The scaling dimensions 
of the operators $U(n,0)$ are $An^2-1$. Operators with greater $n$ are less 
relevant. Since the contribution contains two different operators, it must be 
at least third order [we treat $U(n,0)$ and $U(-n,0)$ as the same operator!]. 
At least one of the two operators must have $|n_i|>1$ (otherwise all operators 
are $U(\pm 1,0)$). Thus, the contribution cannot exceed
$U^2(1,0;E=V)U(2,0;E=V)\sim U^3 V^{6A-2}$. The comparison with 
$I_3\sim U^3 V^{2A+2B-2C-2}$ at $E_F>V>V^*$, yields:
\begin{equation}
	\label{5b_add}
	B<2A+C.
\end{equation}
Note that the above condition is stronger than (\ref{3_add}).

3.3) This case is easy: the contribution must be at least third order again.
Both operators $U(n_i,m_i)$ are less relevant than $U(1,0)$ and $U(0,-1)$ and 
no more relevant than $U(1,-1)$. 
Thus, the contribution is automatically smaller than $I_3$ at any energy scale 
$E_F>V>V^*$.

We now have a full set of conditions at which the third order contribution 
dominates at $V\sim V^*$ and the spin rectification current scales as a 
negative power of the voltage. These are equations 
(\ref{1_add},\ref{0_add},\ref{2_add},\ref{-1_add},\ref{4b_add},\ref{5b_add}).

The above analysis shows that $I_3$ exceeds any contribution to the spin 
rectification current which does not contain $VU_1$ in the whole region 
$E_F>V>V^*$. $I_2$ dominates the remaining contributions for any $V>V^*$. 
The contributions become equal, $I_2=I_3$, at $V=V^{**}=U^{1/[2+2C-2B]}$. 
In the interval of voltages $V^{**}>V>V^*$, the spin rectification current 
is dominated by $I_3$. At $V>V^{**}$, the spin and charge rectification 
currents are dominated by $I_2$.

\subsection{Numerical estimates}

In order to get a feeling about the magnitude of the effect, let us consider 
a particular choice of parameters
$A=B=7/12$, $C=7/24$, $eV\sim 0.01E_F$, $eV^*\sim 10^{-4}E_F$. For such 
$A$, $B$ and $C$ the scaling dimensions of the three most relevant operators 
are the same. The inequalities 
(\ref{0_add},\ref{2_add},\ref{-1_add},\ref{4b_add},\ref{5b_add}) 
are satisfied. The equality $A=B=2C$ corresponds to a limiting case 
of (\ref{1_add}).
One finds that $U\sim 0.01E_F$ and $eV^{**}\sim 0.1E_F$.
Repeating the arguments of the previous section one can estimate the leading 
correction to $I_3$ as $\delta I\sim (eV/E_F)^{7/12}I_3\ll I_3$. The spin 
rectification current is the difference of two opposite electric currents 
of the spin-up and -down electrons times $\hbar/[2e]$.
Even if $E_F$ is as low as $\sim 0.1$ meV, this still corresponds to the 
voltage $V$ of the order of microvolts and the currents\cite{foot} 
(\ref{eq:spin_add}) of spin-up and -down electrons of the order of 
picoamperes, i.e. within the ranges probed in experiments with semiconductor 
heterostructures. Certainly, the current increases, if $E_F$ or $V^*$ is 
increased.


\section{Explicit evaluation of the third order currents}
\label{sec:3rd_order}

The charge or spin currents
in the third order in the potentials $U$ 
are evaluated from the following perturbative
expression:
\begin{multline}
	I_{c,s}^{bs (3)}(V)
	= \frac{(-i)^3}{2!} \sum (n_\uparrow \pm n_\downarrow)
	\int_{C_K} \frac{\mathrm{d}t_1 \mathrm{d}t_2}{\hbar^2} \ 
	\\
	\times
	\langle T_c
	\hat{U}(n_\uparrow,n_\downarrow; 0) 
	\hat{U}(m_\uparrow,m_\downarrow; t_1) 
	\hat{U}(l_\uparrow,l_\downarrow; t_2)
	\rangle, 
\end{multline}
where the sum runs over indices satisfying $n_\sigma + m_\sigma + l_\sigma = 0$
for $\sigma = \uparrow, \downarrow$, $C_K$ is the Keldysh contour 
$-\infty \to 0 \to \infty$, $T_c$ the time order on $C_K$, and we omitted
a constant prefactor.
The operators $\hat{U}$ are given by
\begin{equation}
	\hat{U}(n_\uparrow,n_\downarrow;t)
	= U(n_\uparrow,n_\downarrow) \mathrm{e}^{i (n_\uparrow+n_\downarrow) t eV/\hbar }
	\mathrm{e}^{i n_\uparrow \phi_\uparrow(t) + i n_\downarrow \phi_\downarrow(t)}.
\end{equation}
The most relevant expressions are those arising from the 
combinations $U(1,0) U(0,-1) U(-1,1)$ and $U(-1,0) U(0,1) U(1,-1)$
(see Appendix~\ref{sec:higher_orders}).
The third order contributions to the current contain correlation functions of the form
\begin{multline} \label{eq:P123}
	P(t_1,t_2,t_3) 
	= \langle T_c 
		\mathrm{e}^{\pm i[ \phi_\uparrow(t_1) - \phi_\downarrow(t_2) - \phi_\uparrow(t_3) + \phi_\downarrow(t_3)] } 
	   \rangle	
\end{multline}
We evaluate the correlation functions within the quadratic model described by 
Eq.~\eqref{eq:H0} and use the relations (\ref{eq:transf}) and 
$\langle \tilde{\phi}_{c,s}(t_1) \tilde{\phi}_{c,s}(t_1) \rangle =
-2 \ln(i(t_1-t_2)/\tau_c+\delta)$, with an infinitesimal $\delta>0$ and $\tau_c \sim \hbar/E_F$
the ultraviolet cutoff time.  
This leads to
\begin{multline}\label{mult_add}
	P(t_1,t_2,t_3) = 
	\bigl(i T_c (t_1-t_3)/\tau_c+\delta\bigr)^{2C-2A}\\
	\times\bigl(i T_c (t_2-t_3)/\tau_c+\delta\bigr)^{2C-2B}
        \bigl(i T_c (t_1-t_2)/\tau_c+\delta\bigr)^{-2C},
\end{multline}
where $T_c (t_i-t_j)=(t_i-t_j)$, if time $t_i$ stays later than $t_j$ on the Keldysh contour,
and otherwise $T_c (t_i-t_j)=(t_j-t_i)$.
The expression (\ref{mult_add}) is independent of the $\pm$ signs in Eq.~\eqref{eq:P123}.

The spin and charge current contributions, proportional to  $U(1,0) U(0,-1) U(-1,1)$, 
are complex conjugate to those proportional to 
$U(-1,0) U(0,1) U(1,-1)= U^*(1,0) U^*(0,-1) U^*(-1,1)$. Thus, it is sufficient to calculate 
only the contributions of the first type. In the case of the charge current, their calculation 
reduces to the calculation of the following two integrals over the Keldysh contour: 
\begin{equation}
	\label{k_1_add}
	\int dt_1 dt_3 P(t_1,0,t_3) \exp(ieVt_1/\hbar)
\end{equation}
and
\begin{equation}
	\label{k_2_add} 
	\int dt_2 dt_3 P(0,t_2,t_3) \exp(-ieVt_2/\hbar).
\end{equation}
One of the times $t_1$ and $t_2$ is zero since the current operator is taken at $t=0$ in 
Eq. (\ref{eq:I_Keldysh}). The two integrals can be evaluated in exactly the same way. We will 
consider only the first integral. We find 8 integration regions. They correspond to $2\times 2=4$ 
possibilities for the branches of the Keldysh contour on which $t_1$ and $t_3$
are located and two possible relations $|t_1|>|t_3|$ or $|t_3|>|t_1|$. In all 8 cases, 
we first integrate over $t_3$.
The integral reduces to the Euler $B$-function. Then we integrate over $t_1$. 
This yields a $\Gamma$-function.
Finally, we obtain Eq. (\ref{eq:charge}).

The spin current contains three contributions proportional to $U(1,0) U(0,-1) U(-1,1)$. 
Two of them reduce to the integrals (\ref{k_1_add}) and (\ref{k_2_add}). 
The third contribution is proportional to
\begin{equation}
	\label{k_3_add}
	\int dt_1dt_2 P(t_1,t_2,0)\exp(ieV[t_1-t_2]/\hbar).
\end{equation}
Again we have eight integration regions determined by the choice of the branches of the 
Keldysh contour and the relations $|t_1|>|t_2|$ and $|t_2|>|t_1|$. In each region it is 
convenient to introduce new integration variables: $\tau=|t_1-t_2|$ and $t={\rm min}(t_1,t_2)$. 
The integration over $t$ reduces to a $B$-function. The integration over $\tau$
produces an additional $\Gamma$-function factor. Finally, one obtains Eq. (\ref{eq:spin}).



\pagebreak

\begin{figure}
\begin{center}
\includegraphics[width=\columnwidth]{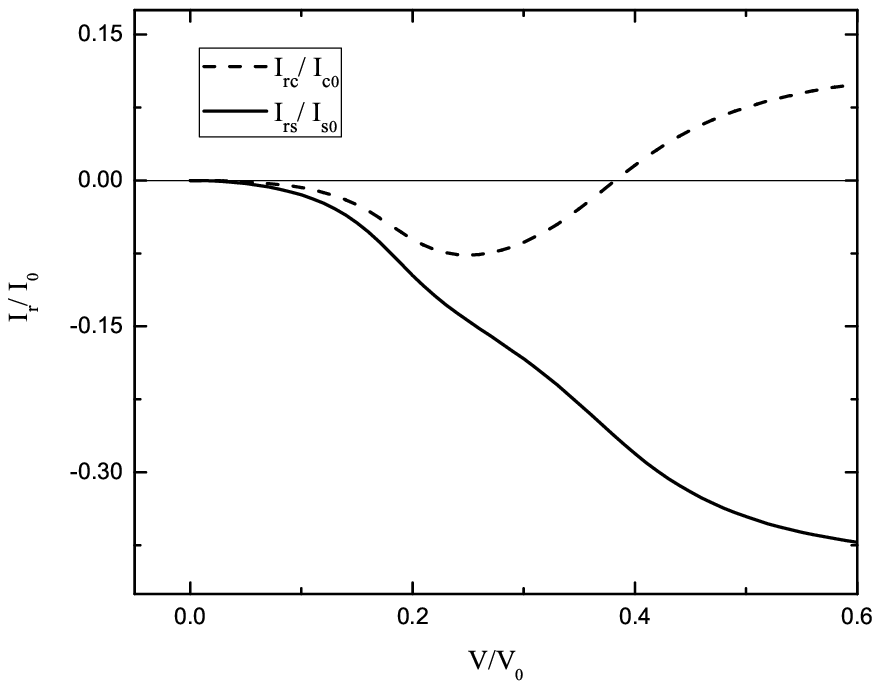}
\caption{Normalized charge rectification
current $I_{c}/I_{c0}$ and spin rectification current $I_{s}/I_{s0}$
versus applied voltage $V/V_0$ for non-interacting electrons with
$E_F=400\epsilon_0$, $\mu H=75\epsilon_0$, $u_{1}=50\epsilon_0a$ and
$u_{2}=-50\epsilon_0a$, where $\epsilon_0=\hbar^2/ma^2$
(see Fig.~\ref{fig:LR} and Appendix~\ref{sec:high_barrier}).
$I_{c0}=50e\epsilon_{0}/\hbar$, $I_{s0}=25\epsilon_{0}$ and
$V_{0}=50\epsilon_{0}/e$ are arbitrary reference currents
and voltage.
\label{fig:noninteract}}
\end{center}
\end{figure}

\end{document}